\documentclass[12pt,fleqn, amsfonts]{article} \textheight=9in
\newcommand{\ra}{\rightarrow}
\newcommand{\bra}{\langle} 
\newcommand{\be}{\begin{equation}}
\newcommand{\ee}{\end{equation}}
\newcommand{\bea}{\begin{eqnarray}}
\newcommand{\eea}{\end{eqnarray}}
\newcommand{\eps}{\varepsilon}

\newcommand{\ffi}{\varphi}

\newcommand{\ep}{\quad {\vrule height 10pt width 10pt depth 2pt}}

\newcommand{\grintl}{[\kern-.18em [}
\newcommand{\grintr}{]\kern-.18em ]}

\newtheorem{lem}{Lemma}[section]
\newtheorem{prop}{Proposition}[section]
\newtheorem{thm}{Theorem}[section]
\newtheorem{cor}{Corollary}[section]

\def\smallR{\hbox{\scriptsize I\kern-.23em{R}}}
\def\0{\hbox{$\mit I$\kern-.70em$\mit O$}}
\def\r{I\kern-.277em R}

\usepackage{amsfonts}

\begin{document}

\title{General Adiabatic Evolution 
with a Gap Condition}

\author{
Alain Joye \\ \\
Institut Fourier
\\ Universit\'e de Grenoble 1,\\ BP 74,
\\38402 St.-Martin d'H\`eres Cedex, \\ France}

\date{  }
\maketitle
\abstract{
We consider the adiabatic regime of two parameters evolution
semigroups generated by linear operators that are analytic in time and
satisfy the following gap condition for all times: the spectrum 
of the generator consists in finitely many isolated eigenvalues of 
finite algebraic multiplicity, away from the rest of the spectrum.
The restriction of the generator to the spectral subspace corresponding to 
the distinguished eigenvalues is not assumed to be diagonalizable.

The presence of eigenilpotents in the spectral decomposition of the
generator forbids the evolution to follow the instantaneous 
eigenprojectors of the generator in the adiabatic limit.
Making use of superadiabatic renormalization, we construct a 
different set of time-dependent projectors, 
close to the instantaneous eigeprojectors of the generator in 
the adiabatic limit, 
and an approximation of the evolution semigroup which intertwines 
exactly between the values of these projectors at the initial and 
final times. Hence, the evolution semigroup follows the constructed 
set of projectors in the adiabatic regime, modulo error terms we control.

}

\section{Introduction}\setcounter{equation}{0}

Singular perturbations of differential equations play an important role
in various areas of mathematics and mathematical physics. Such perturbations
typically  appear when one considers problems that display several different
time and/or length scales. In particular, the semiclassical analysis
of quantum phenomena and the study of evolution equations 
in the adiabatic regime lead to singularly perturbed linear differential
equations which are the object of many recent works. See for example the monographs
 \cite{fed}, \cite{bornemann}, \cite{disj}, \cite{martinez}, \cite{te}. 
The description of certain non conservative phenomena with distinct time scales 
also gives rise to non-autonoumous linear evolution equations,
which are more general than those stemming from conservative systems,
and whose adiabatic regime is of physical relevance, see e.g. \cite{nr}, 
\cite{putterman}, \cite{th}, \cite{lidar}, \cite{lidarprl},  \cite{lidar2}, 
\cite{a-sthesis}, \cite{a-sf0}, \cite{a-s}.

The present paper is devoted to the study of general linear evolution equations 
in the adiabatic limit under some mild spectral conditions on the 
generator. The chosen set up is sufficiently general to cover most applications 
where the time dependent generator is characterized by a gap condition on its spectrum.
Let us describe informally our result, the precise Theorem being formulated in Section 
\ref{mt} below.

We consider a general linear evolution equation in a
Banach space $\mathcal B$ of the form
\be\label{glee}
i\eps \partial_t U(t,s)=H(t)U(t,s), \ \ \ 
U(s,s)=\mathbb I, \ \ s\leq t 
\in [0,1]
\ee 
in the adiabatic limit $\eps \ra 0^+$, for a time-dependent generator
$H(t)$. This equation describes a rescaled non-autonomous evolution generated 
by a slowly varying linear operator $H(t)$. The evolution 
operator $U(t,s)$ evidently depends on $\eps$, even though this is not emphasized 
in the notation.

The generator $H(t)$ is assumed to depend analytically on time and
to have for any fixed $t$ a spectrum $\sigma(H(t))$ divided into two
disjoint parts, $\sigma(H(t))=\sigma(t)\cup \sigma_0(t)$, where
$\sigma(t)$ consists in a finite number of complex eigenvalues 
$\sigma(t)=\{\lambda_1(t), \lambda_2(t), \cdots, \lambda_n(t)\}$
which remain isolated from one another as $t$ varies in $[0,1]$. 
Moreover, the spectral projector of $H(t)$ associated with $\sigma(t)$, denoted
by $P(t)$, is assumed to be finite dimensional. The part of $H(t)$ which corresponds 
to the spectral projector $P_0(t)$ associated with $\sigma_0(t)$ can be unbounded,
bounded  or zero. In the first case we need to assume $H(t)$ generates a {\em bona fide}
evolution operator.

This spectral assumption, or gap condition,  is familiar in the quantum adiabatic context
where $\mathcal B$ is a Hilbert space on which $H(t)$ is further assumed 
to be self-adjoint,  see \cite{bf}, \cite{k0}, \cite{n0}, \cite{asy}, \cite{a-s}, 
for example.  Note that it is still possible to study the quantum adiabatic limit 
by altering the gap condition in different ways, as shown in \cite{ahs}, 
\cite{j0}, \cite{bornemann}, \cite{ae}, \cite{gjjm}, \cite{te0}, \cite{a-sf0}, \cite{a-sf}.

By contrast to previous studies of similar general problems \cite{bn}, \cite{nr}, \cite{mn}, 
\cite{jp2}, \cite{a-s}, we do not assume that the restriction 
of $H(t)$ to the spectral subspace $P(t)\mathcal B$ is diagonalizable. Such situations
take place in the study of open quantum systems by means of phenomenological 
time-dependent master equations,  \cite{lidar},  \cite{lidarprl},  \cite{th}, \cite{lidar2}.
We come back to the approach of \cite{lidar} below.

Therefore, for the part $H(t)P(t)$  of the generator, we have a complete spectral
decomposition
\be\label{sd}
H(t)P(t)=\sum_{j=1}^n \lambda_j(t) P_j(t) + D_j(t),
\ee
where the $P_j(t)$'s are eigenprojectors and the $D_j(t)$'s are 
eigenilpotents associated to the eigenvalue $\lambda_j(t)$ that satisfy
\be\label{propeig}
\sum_{j=1}^nP_j(t)=P(t), \ \  P_j(t)P_k(t)=\delta_{jk}P_j(t), \ \ \mbox{and}\ 
D_j(t)=P_j(t)D_j(t)P_j(t).
\ee 

In case $\mathcal B$ is a Hilbert space on which $H(t)$ is self-adjoint
or if $H(t)$ is diagonalizable with real simple isolated eigenvalues only, the evolution 
$U(t,s)$ {\em follows the instantaneous eigenprojectors} $P_j(t)$ 
in the adiabatic regime in the sense that 
\be\label{stab}
U(t,s)P_j(s)=P_j(t)U(t,s)+O(\eps), \ \ \ \mbox{as }\ \ \eps\ra 0,
\ee
as shown in  \cite{bf}, \cite{k0}, \cite{n0}, \cite{asy}, \cite{a-s}, and
\cite{bn}, \cite{mn}, \cite{jp2}, for example. In other words, transitions
between different spectral subspaces are suppressed as $\eps\ra 0$. This 
relation remains true for certain eigenprojectors if the eigenvalues are allowed 
to have negative imaginary parts, \cite{nr}, \cite{a-s}.  
This fact is also well-known and crucial in the study of the Stokes phenomenon 
appearing in singularly perturbed differential equations
\cite{fed}: under analyticity assumptions, one considers certain paths in the complex 
$t$-plane, called canonical of dissipative paths, along which an equivalent of (\ref{stab}) 
is true in order to get bounds on, or
to compute exponentially small quantities in $1/\eps$ stemming from singularities in the 
complex $t$-plane. Such methods are used  in \cite{jkp}, \cite{jp0}, \cite{jp2} and \cite{j}, to
bound or to compute exponentially small transitions in the adiabatic limit 
when the relevant eigenvalues are real on the real axis.

However, when eigenilpotent are present in the spectral decomposition (\ref{sd}), the 
relation (\ref{stab}) cannot hold in general, even for real valued eigenvalues 
$\lambda_j(t)$. Indeed, the transitions between spectral subspaces  
$P_j(t)U(t,0)P_k(0)$ are typically {\em  exponentially increasing} as $\eps\ra 0$,
rather than vanishing as $\eps$. An explicit example of this fact is provided at 
the end of the Introduction. We come back to this mechanism below.
 
In this context, our main result reads as follows. We construct a 
different set of time-dependent projectors $P_j^{q^*(\eps)}(t)$ 
which  approximates the eigenprojectors $P_j(t)$ in the adiabatic regime $\eps\ra 0$. 
And we show that the evolution $U(t,s)$ can be approximated by a simpler evolution, 
$V^{q^*(\eps)}(t,s)$, which exactly follows the constructed approximations  
$P_j^{q^*(\eps)}(t)$ of the instantaneous eigenprojectors. In other words, we restore the
expected adiabatic behaviour by trading the instantaneous eigenprojectors for
other nearby projectors  in the limit $\eps\ra 0$. Note that since the eigenvalues 
need not be  real in general, we also have to take into account the contributions 
stemming from the
``dynamical phases'' $e^{-i\int_s^t\lambda_j(u)\, du/\eps}$ which can be exponentially
increasing or decreasing as $\eps\ra 0$. In case $H(t)$ is unbounded, we  assume the part 
$H(t)P_0(t)$ generates a semigroup bounded by $|e^{-i\int_s^t\lambda_0(u)\, du/\eps}|$, 
for some function $\lambda_0(t)$.
 \\

 More precisely,  for all $j=1, \cdots, n$ and 
for any  $0\leq t \leq 1$, we construct perturbatively
a set of projectors close to the spectral projectors of $H(t)$, see Section \ref{is},
\be
P_j^{q^\star(\eps)}(t)=P_j(t)+O(\eps), \ \ \mbox{and}\  P_0^{q^\star(\eps)}(t)\equiv 
\mathbb I -\sum_{j=1}^n P_j^{q^\star(\eps)}(t).
\ee 
Let $W^{q^\star(\eps)}(t)$ be the intertwining operator 
naturally associated with the projectors $P_k^{q^\star(\eps)}(t)$, $k=0,\cdots, n$ 
introduced by Kato \cite{k0}, such that
\be\label{w}
W^{q^\star(\eps)}(t)P_k^{q^\star(\eps)}(0)=P_k^{q^\star(\eps)}(t)W^{q^\star(\eps)}(t), 
\ \ \ k=0,\cdots, n.
\ee
The approximation is then defined by 
\be\label{decv}
V^{q^\star(\eps)}(t,0)=W^{q^\star(\eps)}(t)\Phi^{q^\star(\eps)}(t,0)
\ee
where $\Phi^{q^\star(\eps)}(t,s)$  commutes  with all the $P_k^{q^\star(\eps)}(0)$, 
$k=0,\cdots, n$ 
for any $t$ and satisfies a certain singularly perturbed 
linear differential equation, see (\ref{eqphij})
below, which describes  the effective evolution {\em within} 
the fixed subspaces $P_k^{q^\star(\eps)}(0)\mathcal B$. Therefore, 
the following exact intertwining relation holds
\be\label{inter}
V^{q^\star(\eps)}(t,0)P_k^{q^\star(\eps)}(0)=P_k^{q^\star(\eps)}(t)V^{q^\star(\eps)}(t,0), 
\ \ \ k=0,\cdots, n.
\ee
Introducing $\omega(t)=\max_{k=0,\cdots, n}\Im \lambda_k(t)$ to control the norm 
of the ``dynamical phases'', we  prove the existence of 
$\kappa>0$ such that for any  $0\leq t \leq 1$
\be\label{appr}
U(t,0)=V^{q^\star(\eps)}(t,0)+O(te^{-\kappa/\eps} e^{\int_0^t\omega(u)\, du/\eps}),
\ee
where  $\|V^{q^\star(\eps)}(t,0)\|=O(e^{\int_0^t\omega(u)\, du/\eps}e^{D/\eps^\beta})$,
for some $D\geq 0$ and $0<\beta<1.$ Note that the leading term is always meaningful
with respect to the exponentially smaller error term. 
\vspace{.4cm}

In case $\mathcal B$ is a Hilbert space and $H(t)$ is self-adjoint, both the evolution 
and its approximation are unitary and $D$ can be chosen equal to zero. The intertwining
identity (\ref{inter}) and (\ref{appr}) show that the transitions
between the different subspaces $P_j^{q^\star(\eps)}(0)\mathcal B$ are
exponentially small in $\eps$, instead of being of order $\eps$ between the
spectral subspaces of $H$. Constructions leading to approximations $V^{q^\star(\eps)}$ 
of this type with exponentially small error term go under the name {\em superadiabtic 
renormalization}, according to the terminology coined by Berry
\cite{berry}, in this quantum adiabatic context. The first general rigorous 
construction of this type appears in \cite{n1}, but we shall use that of \cite{jp1}.
The statement (\ref{appr}) is thus very similar to the Adiabatic
Theorem of quantum mechanics  \cite{k0}, \cite{n0}, \cite{asy}, \cite{a-s}...
and, more precisely, to the subsequent exponentially accurate versions 
in an analytic context provided in \cite{jp0}, \cite{n1}, \cite{jp1}, \cite{iamp},
\cite{hj}, \cite{hjbham}... or variants therof. 
However, while the improvement of the error term in (\ref{appr}) from 
$O(\eps)$ to $O(e^{-\kappa/\eps})$ by considering 
$P_j^{q^\star(\eps)}$ in place of $P_j$ in the adiabatic context is just that,
{\em improvement}, in case there are non-zero nilpotents in the decomposition 
(\ref{sd}), it becomes {\em necessary} to consider $P_j^{q^\star(\eps)}$ 
and achieve exponential accuracy to get
a result. 

This can be understood as follows.
As $\Phi_\eps(t,0)$ commutes  with all the $P_k^{q^\star(\eps)}(0)$, $k=0,\cdots, n$ 
for any $t$ we can write
\be
\Phi^{q^\star(\eps)}(t,0)=\sum_{k=0}^nP_k^{q^\star(\eps)}(0)\Phi^{q^\star(\eps)}(t,0)
P_k^{q^\star(\eps)}(0)\equiv
\sum_{k=0}^n\Phi^{q^\star(\eps)}_k(t,0).
\ee
The operator $\Phi_j^{q^\star(\eps)}(t,0)$  describing the evolution within 
the fixed subspaces $P_j^{q^\star(\eps)}(0)\mathcal B$  satisfies for $j\geq 1$,
\bea
&&i\eps  \partial_t \Phi_j^{q^\star(\eps)}(t,0)
=(\lambda_j(t)P_j^{q^\star(\eps)}(0)+\widetilde D_j(t, \eps)+O(\eps))
\Phi_j^{q^\star(\eps)}(t,0), \nonumber \\ 
&& \Phi_j^{q^\star(\eps)}(0,0)=P_j^{q^\star(\eps)}(0),
\eea
where $\widetilde D_j(t, \eps)$ denotes the nilpotent 
$\widetilde D_j(t, \eps)={W^{q^\star(\eps)}}^{-1}(t) D_j(t)W^{q^\star(\eps)}(t)$.
We can write 
\be
\Phi_j^{q^\star(\eps)}(t,0)=e^{-\frac{i}{\eps}\int_0^t\lambda_j(u)du}\Psi_j^{q^\star(\eps)}
(t,0),
\ee
where the operator $\Psi_j^{q^\star(\eps)}$ is essentially generated by a nilpotent.
Such  adiabatic evolutions generated by perturbations of 
analytic nilpotents are studied in Section \ref{ng}. We show that $\Psi_j^{q^\star(\eps)}$, 
typically grow when
$\eps\ra 0$ as
\be\label{gro}
\Psi_j^{q^\star(\eps)}(t,0)\simeq e^{c/\eps^{\beta_j}},\ \  \mbox{with}\ \ 0<\beta_j<1,
\ee
wheras $\Psi_j^{q^\star(\eps)}(t,0)$ remains bounded as $\eps\ra 0$ iff 
$ D_j(t)\equiv 0$. The  growth in $e^{1/\beta_j}$ of adiabatic evolutions 
generated by certain nilpotents is already present the works  
\cite{tu} and \cite{s}. 
Hence, to compensate the exponential growth in $1/\eps^{\beta_j}$ of the 
$\Psi_j^{q^\star(\eps)}(t,0)$'s which induces transitions between the 
{\em instantaneous eigenspaces} of the same order, see the example below, 
it is necessary to push the estimates to exponential order, see (\ref{appr}),
by trading the $P_j$'s for the $P_j^{q^\star(\eps)}$.
This requires analyticity of the data, see Section \ref{is}. Analyticity is 
also essential in Section \ref{ng} where the properties of nilpotent generators 
and the adiabatic evolutions they generate are studied.\\

Let us finally comment on the paper \cite{lidar}. It addresses, at a theoretical physics 
level, the  evolution of master equations describing open quantum systems  
in which the components of the Lindblad generator are slowly varying functions 
of time.  Mathematically, this corresponds to a particular case of
problem (\ref{glee}) with a generator containing nilpotents in its decomposition (\ref{sd}). 
The authors argue  under certain {\em implicit} conditions on the evolution, that  it is 
possible to approximate $U(t,0)$ by some operator $V^{\eps}(t,0)$ which
satisfies the intertwining relation (\ref{inter}) with the 
{\em instantaneous projectors} $P_j(t)$ in place of the approximate projectors 
$P_j^{q^\star(\eps)}(t)=P_j(t)+O(\eps)$. However, as we prove,
such a statement cannot be true in general. It does hold, however,
under the hypotheses of \cite{a-s}, that is when the nilpotent 
part of the generator in the corresponding subspace $P_j(t)\mathcal B$ is absent,
together with an {\em a priori} bound on the evolution (see also  remark iii) at the
end of the Section). Or, when the considered
spectral subspace $P_j(t)\mathcal B$ is always decoupled from the others, 
otherwise the error term becomes too large due to the growth (\ref{gro}). 
An example of this sort is 
indeed provided in \cite{lidar}.

\vspace{.4cm}

The paper is organized as follows. We close the introduction by the example
alluded to above and then provide the precise hypotheses and the mathematical 
statement corresponding to our main result. The rest of the paper is devoted 
to the proof of it. The main steps consists in  Section \ref{ng} which 
studies adiabatic evolutions generated by 
(perturbations of) analytic nilpotents. The iterative scheme providing 
the adiabatic renormalization of \cite{jp1} is shortly recalled in Section \ref{is}.
The approximations and its properties are presented in Section \ref{a}.

\vspace{.4cm}

{\bf Acknowledgement:} 
The author  would like to thank C. Ogabi, R. Rebolledo and D. Spehner for useful discussions.

\subsection{About the effect of nilpotents}

We consider here an explicitely solvable model defined by simple generator with two real 
valued distinct eigenvalues possessing a nilpotent in its spectral decomposition. 
We show  that this nilpotent induces exponentially increasing
transitions (in $1/\eps^\beta$, $\beta<1$) between the instantaneous 
eigenspaces, thereby underlying the necessity to use superadiabatic 
renormalization to achieve our result. We also identify the approximated
projectors $P_j^{q^*(\eps)}$ that the evolution follows.

\vspace{.3cm}

Let $H$ be a constant $3\times 3$ matrix in canonical basis 
$\{e_1, e_2, e_3\}$ 
defined by
\be
H=\pmatrix{0 & a & 0\cr 0 & 0 & 0 \cr 0 & 0 & 1}
\ee
and let $L$ be another constant $3\times 3$ matrix defined by
\be\label{l}
L= \pmatrix{0 & 0 & -k\cr -k & 0 & 0 \cr 0 & 0 & 0},
\ee
where the non-zero scalars $a, k$ will be chosen later on.
We set 
\be
S(t):=e^{-it L}, \ \ \ H(t):=S(t)HS^{-1}(t),
\ee
and consider the adiabatic evolution $U(t,0)$ defined for any $t\in [0,1]$ by
\be
i\eps U'(t,0)=H(t)U(t,0), \ \ \ U(0,0)=\mathbb I.
\ee
The spectrum of $H(t)$ is $\{ 0, 1\}$ and its decomposition reads
\be
H(t)=S(t)(0 P_0 + D_0 + 1 P_1 )  S^{-1}(t)\equiv 0 P_0(t)+D_0(t)+1 P_1(t).
\ee
where $P_0=e_1 \bra e_1 |+ e_2  \bra e_2 |$, $P_1= e_3  \bra e_3 |$
and $D_0=a\ e_1  \bra e_2 |$. Here $\{ \bra e_j | \}_{j=1,2,3} $ denotes the 
adjoint basis of $\{e_j \}_{j=1,2,3}$.

The operator $\Omega(t):=S^{-1}(t)U(t,0)$ satisfies 
\be
i\eps \Omega'(t)=(H-\eps L)\Omega(t), \ \ \ \Rightarrow 
\Omega(t)=e^{-it(H-\eps L)/\eps}.
\ee
The matrix  $H-\eps L$ is now diagonalizable and its spectrum is 
\be
\{1, +\sqrt{\eps a k}, -\sqrt{\eps a k} \}\equiv \{1, \lambda_+(\eps), 
-\lambda_+(\eps)\}\equiv \{1, \lambda_+(\eps), 
\lambda_-(\eps)\},
\ee 
where $\sqrt{\cdot}$ denotes
any branch of the square root function. The corresponding spectral projectors
are denoted by $P_1(\eps), P_+(\eps)$ and $P_-(\eps) $ and they are given by
\bea
&&P_1(\eps)=\pmatrix{0 & 0 & \frac{\eps k}{1-\eps ak}\cr 0 & 0 & 
\frac{\eps^2 k^2}{1-\eps ak} \cr 0 & 0 & 1}, \\ &&P_+(\eps)=
\pmatrix{\frac{\lambda_+(\eps)}{\lambda_+(\eps)-\lambda_-(\eps)} & 
\frac{a}{\lambda_+(\eps)-\lambda_-(\eps)} & \frac{\lambda_+(\eps)\eps k}
{(\lambda_+(\eps)-\lambda_-(\eps))(\lambda_+(\eps)-1)}\cr \frac{\eps k}
{\lambda_+(\eps)-\lambda_-(\eps)} & \frac{\lambda_+(\eps)}
{\lambda_+(\eps)-\lambda_-(\eps)} & 
\frac{\eps^2 k^2}{(\lambda_+(\eps)-\lambda_-(\eps))(\lambda_+(\eps)-1)} 
\cr 0 & 0 & 0},
\eea
and $P_-(\eps)$ has the same expression as $P_+(\eps)$ with indices $+ $ 
and $-$ exchanged. Note that $P_\pm(\eps)\simeq \pm a/\sqrt{\eps ak}$ as
$\eps\ra 0$. 
Whereas the projectors 
\bea\label{plg}
P_1(\eps)&=&P_1+O(\eps)\\
P_0(\eps)&=& P_+(\eps)+P_-(\eps)=P_0+O(\eps)
\eea
admit expansions in powers of $\eps$.
Hence
\be\label{expu}
U(t,0)=S(t)(e^{-it/\eps}P_1(\eps)+e^{-it\lambda_+(\eps)/\eps}P_+(\eps)
+e^{-it\lambda_-(\eps)/\eps}P_-(\eps)),
\ee
so that, as $\eps\ra 0$,
\be
\|U(t,0)\|\simeq \frac{a}{2\sqrt{\eps ak}}(e^{-it\sqrt{\eps ak}/\eps}-e^{it\sqrt{\eps ak}/\eps})
\ee
which diverges, whatever the nonzero value of $ak$ is. \vspace{.3cm}

We now choose $ak <0 $ and $\lambda_\pm(\eps)=\pm i\sqrt{\eps |ak|}\in i\mathbb R$, 
for definiteness.
Since $P_k(t)U(t,0)P_j(0)=S(t)P_k\Omega(t)P_j$, $j, k\in\{0,1\}$,
where $S(t)$ is independent of $\eps$, it is enough to compute
$P_k\Omega(t)P_j$ to get the behaviour in $\eps$ of the transitions
between the corresponding instantaneous subspaces.  We get for $t>0$
and $P_1=e_3 \bra e_3|$,
\bea\label{exptra}
P_0\Omega(t)P_1&=&
e^{t\sqrt{|ak|}/\sqrt{\eps}}\pmatrix{-\frac{\eps k}{2}\cr
\frac{i\eps^{3/2} k^2}{2\sqrt{|ak|}}\cr 0 }\bra e_3|+ O(\eps), \ \ \mbox{as} \ \ 
\eps\ra 0\\
P_1\Omega(t)P_0&\equiv& 0 \label{sym}
\eea
The first formula thus implies that the evolution $U(t,0)$ does not follow the  
instantaneous eigenprojector $P_1(t)$, whereas the second formula simply 
reflects the non-generic fact that $P_0$ is invariant under $H-\eps L$ in 
our example, see the remarks below.

The model being explicitly solvable, we can readily identify the approximated
projectors the evolution follows. Setting for $j=0,1$
\bea\label{ppp}
P_j^*(t,\eps):=S(t)P_j(\eps)S^{-1}(t)=P_j(t)+O(\eps),
\eea
we compute by means of (\ref{expu}) and (\ref{plg})
\be
U(t,0)P_j^*(0,\eps)=P_j^*(t,\eps)U(t,0).
\ee
Thus the evolution $U(t,0)$ exactly follows the projectors (\ref{ppp}) 
whereas the transition from $P_1(0)$ to $P_0(t)$ are exponentially 
large in $1/\sqrt{\eps}$.

\vspace{.4cm}

\noindent {\bf Remarks:}\\ 
i) If the product $ak\in\mathbb C\setminus \mathbb R^+$, 
a similar result holds. We took $ak<0$ for simplicity.
If the product $ak$ is positive, the transition does vanish in the
limit $\eps\ra 0$. This is due to the fact that the 
spectral projector $P_0$ corresponding to the unperturbed eigenvalue
$0$ of $H$ is of dimension 2.  For the natural generalization of this example 
with $\dim P_0=d$, $d>2$, the following holds. Generically, the splitting of 
the unique eigenvalue zero of the nilpotent $P_0H$ by a perturbation of order $\eps$ 
yields $d$ perturbed eigenvalues $\lambda_j(\eps)\simeq \alpha \eps^{1/d}e^{j 2i\pi/d}$, 
$j=0,\cdots, d-1$, $\alpha\in\mathbb C$, see \cite{k}.  Hence, one of them  has a non vanishing 
imaginary part that produces exponentially growing contributions as $\eps\ra 0$. \\
ii) As already mentioned, this example is non-generic in the sense that $P_0$ is 
invariant under $\Omega(t)$, see (\ref{sym}). The choice of non-generic $L$ 
(\ref{l}) was made to keep the formulas simple. However, as should be clear from 
the analysis, a generic choice for $L$ implies an exponential increase as 
$\eps\ra 0$ for both  $P_1\Omega(t)P_0$ and $P_0\Omega(t)P_1$, when 
$ak\in\mathbb C\setminus \mathbb R^+$.\\
iii) The real unperturbed eigenvalues $0$ and $1$ can be replaced by any different 
complex numbers $\lambda_0$ and $\lambda_1$ without difficulty. The main
consequence is that the exponents in (\ref{expu}) have to be changed according to
$\lambda_\pm(\eps)\mapsto \lambda_0+\lambda_\pm(\eps)$ and $1\mapsto \lambda_1$. 
One can assume without loss that $\Im \lambda_j\leq 0$, $j=0,1$.
Observe that if $\lambda_0$ is real and $\Im \lambda_1<0$,  conclusions similar
to (\ref{exptra}) can be drawn. In case $\Im \lambda_0<0$ and $\lambda_1$ is real, 
the transition   $P_0\Omega(t)P_1$ is of order $\eps$, $\eps\ra 0$. This is 
a case where the results of \cite{a-s} apply, since the evolution 
(\ref{expu}) becomes uniformly bounded in $\eps$ due to the exponential decay
stemming from $\Im \lambda_0<0$.

\section{Main Result}\label{mt}\setcounter{equation}{0}

Let us specify here our hypotheses and state our result.

Let $a>0$ and $S_a=\{z\in \mathbb C \ | \ \mbox{dist}(z,[0,1])< a\}$.

{\bf H1:} \\
Let $\{H(z)\}_{z\in \bar{S}_a}$ be a family of closed operators densely
defined on a common domain $\mathcal D$ of a Banach space $\mathcal B$
and for any $\ffi \in \mathcal D$, the map $z\mapsto H(z)\ffi$ 
is analytic in $S_a$.
\vspace{.5cm}

As a consequence, the resolvent 
$R(z,\lambda)=(H(z)-\lambda)^{-1}$ is locally analytic in $z$ for 
$\lambda \in \rho(H(z))$, where  $\rho(H(z))$ denotes the resolvent set
of $H(z)$.
\vspace{.5cm}

{\bf H2:} \\
For $t\in [0,1]$, the spectrum of $H(t)$ is of the form
$\sigma(H(t))=\sigma(t)\cup \sigma_0(t)$, and there exists $G>0$ such that
$$\inf_{t\in [0,1]}\mbox{dist}(\sigma(t),\sigma_0(t))\geq G.$$  
Moreover,  
$\sigma(t)=\{\lambda_1(t), \lambda_2(t), \cdots, \lambda_n(t)\}$
where $ \lambda_j(t)$, $j=1,\cdots, n$, $n<\infty$, are eigenvalues of 
constant multiplicity $m_j<\infty$ such that 
$$\inf_{t\in [0,1]\atop j\neq k}\mbox{dist}( \lambda_j(t), \lambda_k(t))
\geq G .$$
\vspace{.5cm}

Let  $\Gamma_j\in \rho(H(t))$ be a loop encircling $\lambda_j(t)$ only.
The finite dimensional spectral projectors corresponding to the eigenvalues
$\lambda_j(t)$ are given by
\be\label{defp}
P_j(t)=-\frac{1}{2\pi i}\int_{\Gamma_j} R(t,\lambda) d\lambda \ \ 
\mbox{and we set } \ \  P_0(t)=\mathbb I-\sum_{j=1}^n P_j(t)
\equiv\mathbb I- P(t) .
\ee
The loop  $\Gamma_j$ can be chosen locally independent of $t$. 
It is a classical perturbative fact, see \cite{k}, that 
{\bf H2} also holds for the spectrum of $H(z)$ with $z\in S_a$, 
provided $a$ is small enough, and that the eigenvalues are analytic
functions in $S_a$. By this we mean that the $\inf_{t\in [0,1]}$ can
be replaced by $\inf_{t\in S_a}$ in {\bf H2}.
Hence, (\ref{defp}) also holds for 
$z\in S_a$ and  $z\mapsto P_k(z)$ is analytic in $S_a$, for $k=0,\dots, n$. 
Consequently, the eigenilpotents given by 
$D_j(z)=(H(z)-\lambda_j(z))P_j(z)$ are analytic in $S_a$ as well.

\vspace{.5cm}

We now state a technical hypothesis needed to deal with evolution 
operators generated by unbounded generators. In case one works with 
bounded operators only, this hypothesis is not necessary.

\vspace{.5cm}

{\bf H3:} \\
Let $H_0(t)=P_0(t)H(t)P_0(t)$. 
There exists a $C^1$ complex valued function $t\mapsto \lambda_0(t)$ 
such that for all $t\in [0,1]$, 
$H_0(t)+\lambda_0(t)$ generates a contraction semigroup and 
$0\in \rho((H_0(t)+\lambda_0(t))$.

\vspace{.5cm}

In  other words, {\bf H3} says that  the solution
$T(s)=e^{-i\lambda_0(t) s}e^{-iH_0(t)s}$ to the strong 
equation on $\mathcal D$ 
$i\partial_s T(s)=(H_0(t)+\lambda_0(t))T(s)$
satisfies  $\|T(s)\|\leq 1$, for all $s\geq 0$. By Hille-Yoshida's Theorem, 
{\bf H3} is equivalent to the following spectral condition for any $t\in [0,1]$, 
\be
[0,\infty)\subset \rho(-iH_0(t)-i\lambda_0(t)) \ \ \ \mbox{and} \ \ \
\|(iH_0(t)+i\lambda_0(t)+\lambda)^{-1}\|\leq \frac{1}{\lambda} , \ \ 
\forall \lambda >0.
\ee

\vspace{.5cm}

This hypothesis implies that the equation
\be\label{evol0}
i\eps \partial_t U_0(t,s)\ffi =H_0(t)U_0(t,s)\ffi, \ \ \ 
U_0(s,s)\ffi= \ffi, \ \ s\leq t 
\in [0,1], \ \ \forall \ffi \in \mathcal D,
\ee
defines a unique {\em strongly continuous two-parameter evolution operator}
$U_0(t,s)$. It means that $U_0(t,s)$ is uniformly bounded, strongly continuous
in the triangle $0\leq s \leq t\leq 1$ and satisfies the
relation $U_0(t,s)U_0(s,r)=U_0(t,r)$ for any $0\leq r\leq s \leq t\leq 1$. 
Moreover, $U_0(t,s)$
maps $\mathcal D$ into $\mathcal D$, also satisfies
\be\label{evol2}
i\eps \partial_s U_0(t,s)\ffi =-U_0(t,s)H_0(s)\ffi, \ \ 
\forall \ffi \in \mathcal D,
\ee
 and is such that
$H_0(t)U_0(t,s)(H_0(s)+\lambda_0)^{-1}$ is bounded and continous in the 
triangle $0\leq s\leq t\leq 1$. 
Moreover, see \cite{rs}, Thm X.70., the following bound holds
\be\label{estu0}
\|U_0(t,s)\|\leq e^{\int_s^t \Im \lambda_0(u) \, du/\eps}, \ \ \forall s\leq t \in [0,1].
\ee

Since $H(t)=H_0(t)+P(t)H(t)P(t)$ where $P(t)H(t)P(t)$ is bounded and 
analytic in $t$, Hypothesis {\bf H3} also implies existence and uniqueness
of a {\em bona fide} evolution operator $U(t,s)$ associated with the equation 
\be\label{evol}
i\eps \partial_t U(t,s)\ffi =H(t)U(t,s)\ffi, \ \ \ 
U(s,s)\ffi= \ffi, \ \ s\leq t 
\in [0,1], \ \ \forall \ffi \in \mathcal D,
\ee 
see \cite{kr}, Thm 3.6, 3.7 and 3.11.

\begin{thm}\label{math}
Assume {\bf H1}, {\bf H2} and {\bf H3} and consider $U(t,0)$ defined by
(\ref{evol}). For $k=0,\cdots, n$, let $P_k^{q^\star(\eps)}(t)=P_k(t)+O(\eps)$
be defined by (\ref{pq}), (\ref{qstar}) and 
$V^{q^\star(\eps)}(t,0)=W^{q^\star(\eps)}(t)\Phi^{q^\star(\eps)}(t,0)$
given by (\ref{vq}), (\ref{kato}), (\ref{defphi}) and (\ref{qstar}). 
Define $\omega(t)=\max_{k=0,\cdots, n}\Im \lambda_{k}(t)$.
Then, there exists a constants $\kappa>0$ such that for any  $0\leq t \leq 1$
\begin{eqnarray*}
&& e^{-\int_0^t \omega(u)\, du/\eps}U(t,0)P_k^{q^*(\eps)}(0)=e^{-\int_0^t \omega(u)\, du/\eps} 
V^{q^*(\eps)}(t,0)P_k^{q^*(\eps)}(0)\\
\nonumber
&&\qquad \qquad  \qquad  \qquad \qquad  \qquad +O(t e^{-\kappa /\eps}\sup_{ 0 \leq s \leq t}\|e^{-\int_0^s \omega(u)\, du/\eps} 
V^{q^*(\eps)}(s,0)P_k^{q^*(\eps)}(0)\|),
\end{eqnarray*}
with
$$
V^{q^\star(\eps)}(t,0)P_k^{q^\star(\eps)}(0)=P_k^{q^\star(\eps)}(t)
V^{q^\star(\eps)}(t,0), \ \ \ k=0,\cdots, n.
$$
Moreover, for all $k\geq 0$ there exists $0\leq \beta_k <1 $, $c_k>0$, 
and $d_k\geq 0$, with $d_0=0$, such that  
$$\|V^{q^\star(\eps)}(t,0)P_k^{q^\star(\eps)}(0)\|
\leq c_k e^{d_k/\eps^{\beta_k}}e^{\Im\int_0^t\lambda_k(u)du/\eps},$$
with $d_j=0$, if and only if $D_j(t)\equiv 0$,  $j\in \{1,\cdots, n\}$, in
(\ref{sd}).
\end{thm}
As a direct 
\begin{cor}
Under the hypotheses of Theorem \ref{math},  there exists  $\kappa>0$,
$0<\beta<1$, and $D\geq 0$ such that
$$
U(t,0)=V^{q^*(\eps)}(t,0)+
O(te^{-\kappa /\eps}e^{\int_0^t \omega(u)\, du/\eps}),
$$
where $V^{q^*(\eps)}(t,0)=O(e^{\int_0^t \omega(u)\, du/\eps}e^{D/\eps^{\beta}})$.
\end{cor}

\noindent
{\bf Remarks:}\\
0) The equivalent results hold if the initial time $0$ is replaced by any 
$0\leq s\leq t$, {\em mutatis mutandis}. See Subsection \ref{eop}.\\
i) As is obvious from the formulation, the natural operators to control
are $e^{-\int_0^t \omega(u)\, du/\eps}U(t,0)$ and $e^{-\int_0^t \omega(u)\, du/\eps} 
V^{q^*(\eps)}(t,0)$. \\
ii) As particular cases of Theorem \ref{math}, we recover the results
of \cite{bn}, \cite{nr}, \cite{mn}, \cite{jp2}, \cite{a-s}.\\
iii)  In case $\kappa$ is sufficiently large, the different components 
of the leading order term have amplitudes whose instantaneous exponential decay or growth 
rates in $1/\eps$ may change with time. More precisely, assume that
\be
\int_0^t \Im \lambda_k(u)\, du > \int_0^t \omega(u)\, du-\kappa, \ \ \forall t\in [0,1], \ \ 
\mbox{and} \ \ \ \forall k=0,\cdots, n.
\ee
This can be achieved by perturbating weakly 
a generator for which all $\lambda_k$ are real valued, for example. Then,
for any initial condition
\be
\ffi=\sum_{k=0}^nP_k^{q^\star(\eps)}(0)\ffi\equiv \sum_{k=0}^n\ffi_k(\eps) \in \mathcal D,
\ee
we get
\be
U(t,0)\ffi=\sum_{k=0}^ne^{-i\int_0^t\lambda_k(u)du/\eps}\Psi_k^{q^\star(\eps)}(t,0)\ffi_k(\eps)
+O(te^{-\kappa /\eps}e^{\int_0^t \omega(u)\, du/\eps}),
\ee
where the error term is exponentially smaller than the leading terms. Each term of the sum
decays or grows as $\eps\ra 0$ with an instantaneous exponential rate given by 
$\Im\int_0^t\lambda_k(u)du/\eps$. Depending on the functions 
$t\mapsto \Im\int_0^t\lambda_k(u)du/\eps$, the index of the component  which is
the most significant may vary with time. \\
iv) In case all  $\lambda_k(t)$ are real, $k=0,\cdots, n$, and $H(t)$ is 
diagonalizable, we can take $d_k=0$ for all $k=0,\cdots, n$, and 
$\omega(t)\equiv 0$. The evolution $U$ and its approximation 
$V^{q^*(\eps)}$ are then uniformly bounded in $\eps$ and differ by an error of order
$e^{-\kappa /\eps}$.  Theorem \ref{math} thus  generalizes 
Thm 2.4 in \cite{jp2} in the sense that
we allow permanently degenerate eigenvalues $\lambda_j(t)$, whereas they were assumed
to be simple in \cite{jp2}.

\section{Preliminary Estimates}\label{pe}\setcounter{equation}{0}

We start by recalling a perturbation formula for evolution operators that we
will use several times in the sequel.

\vspace{.3cm}

Let $\{A(t)\}_{t\in [0,1]}$ be a densely defined family of linear operators 
on a common domain 
$\mathcal D$ of a Banach space $\mathcal B$, and assume 
$t\mapsto A(t)$ is strongly continuous. Let $B(t)$ be linear, bounded and strongly
continuous in $t\in [0,1]$. Assume there exist two-parameter evolution
operators $T(t,s)$ and $S(t,s)$ associated with the equations
\bea
i\partial_t T(t,s)\ffi &=&A(t)T(t,s)\ffi , \ \ \ T(s,s)=\mathbb I, \\
i\partial_t S(t,s)\ffi &=&(A(t)+B(t))S(t,s)\ffi , \ \ \ S(s,s)=\mathbb I, 
\eea
for  all $\ffi\in \mathcal D \ \ \mbox{and} \ s\leq t\in [0,1]$.
Then, for any $\ffi\in \mathcal D$, and any $r\leq s \leq t \in [0,1]$,
\be
i\partial_s (T(t,s)S(s,r)\ffi) = T(t,s)B(s)S(s,r)\ffi, 
\ee
so that by integration on $s$ between $r$ and $t$, 
\be\label{difst}
S(t,r)\ffi=T(t,r)\ffi-i\int_r^t ds T(t,s)B(s)S(s,r)\ffi. 
\ee
Iterating this formula, we deduce the representation
\bea\label{dyson}
&&S(t,r)=\sum_{n\geq 0}(-i)^n \int_r^t ds_1 \int_r^{s_1} ds_2 \cdots 
\int_r^{s_{n-1}} ds_n \nonumber\\
 &&\qquad \qquad\qquad \qquad \times \quad T(t,s_1)B(s_1)T(s_1, s_2)
B(s_2)\cdots B(s_n)
T(s_n, r).
\eea
Further assuming that $T(t,s)$ satisfies the bound 
\be
\| T(t,s)\| \leq Me^{\int_s^t \omega(u) du},  
\ee
for a constant $M$ and a real valued integrable function $u\mapsto \omega(u)$, we get
from (\ref{dyson})
\be\label{pert}
\| S(t,s)\| \leq M e^{\int_r^t (\omega(u) + M\|B(u)\|)\, du}.
\ee
As a first application of (\ref{pert}), we get from (\ref{estu0})
a first estimate on $U(t,s)$ that we will improve later on
\be
\|U(t,s)\|\leq e^{\int_s^t (\Im \lambda_0(u)+\|P(u)H(u)P(u)\| )\, du/\eps}.
\ee

\section{Nilpotent Generators}\label{ng}\setcounter{equation}{0}

For later purposes, we study here the adiabatic evolution generated by
an analytic nilpotent, in a finite dimensional space. We assume\\

{\bf N1:}\\
For any $z\in S_a$,  $N(z)$ is an analytic nilpotent valued 
operator in a linear space $\cal B$ of finite dimension such that 
for a fixed integer $d\geq 0$, $N(z)^d\equiv 0$.\\

\vspace{.4cm}

The detailed analysis of the properties of analytic nilpotent
matrices  is performed in Section 5 of the book \cite{baum}. 
It is shown in particular that such operators have the following
structure. 
For any nilpotent 
$N(z)$ satisfying {\bf N1} in $S_a$, there exists a finite set 
of points $Z_0\subset S_{a'}$, with $a'<a$, and,  there exists a family 
of invertible operators $\{ S(z)\}_{z\in S_{a'}\setminus Z_0}$ such that
 for any 
$z\in S_{a'}\setminus Z_0$,
\be
N(z)=S^{-1}(z)NS(z)
\ee
with $S(z)$ and $S^{-1}(z)$ meromorphic in $S_{a'}$ and
regular in  $S_{a'}\setminus Z_0$. The set $Z_0$ where
$N(z)$ is not similar to the constant nilpotent $N$ is 
called the set of
{\em weakly splitting points} of $N(z)$. At these points, the range
and kernel of $N(z)$ change.

\vspace{.4cm}

We consider $Y(t,s)$, defined as the solution to 
\be\label{evonil}
\eps \partial_t Y(t,s)=N(t)Y(t,s), \ \ Y(s,s)=\mathbb I, \ \
\forall s, t \in [0,1], 
\ee
and estimate the way $Y(t,s)$ depends on $\eps$, as $\eps\ra 0$. Note
that we don't need to impose $s\leq t$ since we deal with
 bounded generators.\\

In case $N$ is constant, with $N^{d-1}\neq 0$, 
$Y(t,s)=e^{(t-s)N/\eps}$ behaves polynomially in $1/\eps$, 
{\em i.e.} like $((t-s)/\eps)^{d-1}$, 
as $\eps\ra 0$. When $N(t)$ is not constant, one may expect that
$Y(t,s)$ explodes less fast than $e^{c/\eps}$, which is the worst
behavious as $\eps\ra 0$ for bounded generators. In such cases, however,
$Y(t,s)$ grows typically faster than polynomially in $1/\eps$, as the
following  example shows. For $N(z)$  given by
\be\label{exn}
N(t)=\pmatrix{t & -1 \cr t^2 & -t},
\ee
we get that the solution $Y(t,0)$ to (\ref{evonil}) reads
\be\label{exe}
Y(t,0)=\pmatrix{\cosh(\frac{t}{\sqrt{\eps}}) & -\frac{1}{\sqrt{\eps}}
\sinh(\frac{t}{\sqrt{\eps}})\cr t\cosh(\frac{t}{\sqrt{\eps}})-
\sqrt{\eps} \sinh(\frac{t}{\sqrt{\eps}})   & \cosh(\frac{t}{\sqrt{\eps}})
 -\frac{t}{\sqrt{\eps}}\sinh(\frac{t}{\sqrt{\eps}}) },
\ee
which behaves as $e^{t/\sqrt{\eps}}$, when $\eps\ra 0$. The growth is
nevertheless slower than exponential in $1/\eps$. We show that
the characteristic behaviour of $Y$ generated by an analytic nilpotent
operator is similar. For later purposes, we actually consider generators
given by an order $\eps$ perturbation of a nilpotent.

\begin{prop}\label{nilpot}
Suppose the nilpotent $N(t)$ satisfies {\bf N1} and let 
$\{ A(t)\}_{t\in [0,1]}$ be a  $C^0$ family of operators on $\mathcal B$.
Then, there exist $c>0$ and $0<\beta<1$ such that 
the solution $Y(t,s)$ of 
\be\label{evonilp}
\eps \partial_t Y(t,s)=(N(t)+\eps A(t))Y(t,s), \ \ Y(s,s)=\mathbb I, \ \
\forall s, t \in [0,1], 
\ee
satisfies uniformly in $t,s\in [0,1]$
$$
\|Y(t,s)\|\leq ce^{c/\eps^\beta}.
$$
\end{prop}
{\bf Remarks:}\\
i) Asymptotic expansions as $\eps\ra 0$ of solutions to such equations
are derived in \cite{tu}, \cite{s},  in the neighbourhood of points which are not 
weakly splitting points for $N(z)$.\\
ii) In case both $0$ and $1$ are not  weakly splitting points, it is possible
to take $\beta=(d-1)/d$, which is the optimal exponent, see the example. 
As we shall not need such improvements, 
we don't give a proof.\\
iii) The adiabatic evolution generated by an analytic nilpotent does 
not have to grow exponentially fast in $1/\eps^{\beta}$, as $\eps\ra 0$. 
Consider for example (\ref{exn}) and 
(\ref{exe}) along the imaginary $t$-axis. However, such evolutions cannot
be uniformly bounded in $\eps$, as the next Lemma shows, under slightly
stronger conditions. \\
iv) It is actually enough to assume $t\mapsto \|A(t)\|$ is uniformly 
bounded on $[0,1]$.

\begin{lem}\label{compic}
Assume $\{ N(t)\}_{t\in [0,1]}$ is a  $C^1$ family of nilpotents and 
$\{ A(t)\}_{t\in [0,1]}$ is a $C^1$ family of operators on $\mathcal B$.
Consider $Y(t,s)$ the solution to (\ref{evonilp}). Then
$$\sup_{\eps >0}\|Y(t,s)\|<\infty \Longleftrightarrow N(u)\equiv 0 \ \  
\forall s\leq u \leq t.$$
\end{lem}

\vspace{.4cm}

\noindent
{\bf Proof of Proposition \ref{nilpot}:}
The proof consists in two steps. First we prove the result for generators
with more structure and then, making use of the results of Section 5 in \cite{baum} 
on the detailed structure of analytic nilpotents, we extend it to the
general case.

\begin{lem}\label{easy}
Assume $N(t)=S^{-1}(t)N S(t)$ where $N$ satisfies $N^d=0$ and where 
$\{S(t)\}_{t\in [0,1]}$ is a $C^1$ family of invertible operators.
Let $\{ A(t)\}_{t\in [0,1]}$ be a  $C^0$ family of operators and set
$B(t)=S(t)A(t)S^{-1}(t)+S'(t)S^{-1}(t)$.
Then, there exists $c>0$ such that the solution $Y(t,s)$  of (\ref{evonilp}) satisfies
$$
\|Y(t,s)\|\leq \|S^{-1}(t)\|\|S(s)\|\frac{c}{\eps^{(d-1)/d}}
e^{ \int_s^t (1+c\|B(u)\|) du /\eps^{(d-1)/d}}, 
\ \ \forall s\leq t \in [0,1].
$$
\end{lem}
{\bf Remarks:}\\
0) The constant $c$ depends on $N$ only.\\
i) If $s\geq t$, the same estimate holds with 
$\int_t^s \|B(u)\| du$ in the exponent.\\
ii) This Lemma also holds in infinite dimension.

\vspace{.3cm}

\noindent
{\bf Proof of Lemma \ref{easy}: }  
Let $Z(t,s)=S(t)Y(t,s)S^{-1}(s)$. This operator satisfies by construction
\be\label{evoz}
\eps \partial_t Z(t,s)=(N+\eps B(t))Z(t,s), \ \ Z(s,s)=\mathbb I, \ \
\forall s, t \in [0,1].
\ee
Let us compare $Z(t,s)$ with 
\be
Z_0(t,s)=e^{N(t-s)/\eps}, \ \  s, t \in [0,1]
\ee
by means of (\ref{dyson}). We get
\bea\label{back}
&&Z(t,r)=\sum_{n\geq 0} \int_r^t ds_1 \int_r^{s_1} ds_2 \cdots 
\int_r^{s_{n-1}} ds_n \nonumber\\
 &&\qquad \qquad\qquad \qquad \times \quad Z_0(t,s_1)B(s_1)Z_0(s_1, s_2)
B(s_2)\cdots B(s_n)
Z_0(s_n, r).
\eea
Consider now 
\be
Z_\delta(s)=e^{(N-\delta)s}, \  \ \mbox{for } \ \  \delta>0. 
\ee
This operator is such that there exists a $c>0$, which depends on 
$N$ only, such that
\be\label{eszd}
\| Z_\delta(s)\|\leq c/\delta^{d-1} \ \ \forall s\geq 0, \ \mbox{and} \ \
0<\delta \leq 1.
\ee
Indeed, on the one hand, 
we have for $s\geq s_0$, with $s_0$ large enough
$
\|Z_\delta(s)\|\leq K e^{-\delta s}s^{d-1},
$
where $K$ is some constant which 
depends on $N$ only.
Maximizing over $s\geq 0$, we get
$
e^{-\delta s}s^{d-1}\leq  e^{1-d}\frac{(d-1)^{d-1}}{\delta^{d-1}}.
$
On the other hand, for all  $0\leq s\leq s_0$, we have
$\|Z_\delta(s)\|\leq e^{s_0\|N\|}$, 
so that if $0<\delta\leq 1$, (\ref{eszd}) holds with 
$c=\max(e^{s_0\|N\|},K ((d-1)/e)^{d-1} )$.

Coming back to (\ref{back}) in which we make use of the relation
\be
Z_0(t,s)=Z_\delta((t-s)/\eps)e^{\delta (t-s)/\eps},
\ee
and (\ref{eszd}), we get
\bea
\|Z(t,r)\|&\leq& e^{\delta (t-r)/\eps}\sum_{n\geq 0} \int_r^t ds_1 
\int_r^{s_1} ds_2 \cdots 
\int_r^{s_{n-1}} ds_n \nonumber\\
 &&\times  \| Z_\delta((t-s_1)/\eps)B(s_1)Z_\delta((s_1-s_2)/\eps)
B(s_2)\cdots B(s_n)
Z_\delta((s_n -r)/\eps)\|\nonumber\\
&\leq& \frac{ce^{\delta (t-r)/\eps}}{\delta^{d-1}} \sum_{n\geq 0} 
\frac{(c \int_r^t \|B(s)\| \, ds /\delta^{d-1} )^n}{n!}\nonumber\\
&=&\frac{ce^{\delta (t-r)/\eps}}{\delta^{d-1}}e^{c \int_r^t \|B(s)\| \, ds /\delta^{d-1}}.
\eea
The left hand side is independent of $\delta$, which we can chose as
$\delta=\eps^{1/d}$, so that we eventually get
\be
\|Z(t,r)\|\leq 
\frac{c}{\eps^{(d-1)/d}}e^{\int_r^t (1 + c\|B(s)\|)  \, ds/\eps^{(d-1)/d}},
\ee 
from which the result follows.
\ep 

\vspace{.3cm}
Let us go on with the proof of the Proposition. 
If $Z_0\cap [0,1]=\emptyset$, Lemma (\ref{easy}) applies and 
Proposition \ref{nilpot} holds. If not, there exist a finite set
of real points $\{0\leq t_1 < t_2 <\cdots < t_m\leq 1 \}$ and a finite
set of integers $\{p_j\}_{j=1, \cdots, m} $ such that
\be\label{ss}
\max (\|S(t)\|, \|S^{-1}(t)\|, \|S'(t)S^{-1}(t)\|)=O (1/(t-t_j)^{p_j})
, \ \ \mbox{as}\ \ \ t\ra t_j.
\ee 
Since $Y$ is 
an evolution operator, we can split the integration range in finitely
many intervals, so that it is enough to control $Y(t,s)$ for  
$s\leq t\in [v,w]\subset \mathbb R$ where $[v,w]$ contains 
one singular point only. Call this singular point $t_0$ and the corresponding
integer $p_0$.

Assume to start with that $v<t_0<w$. Let $\delta>0$ be small
enough and $v\leq s<t_0<t\leq w$  so that we can write
\be
Y(t,s)=Y(t,t_0+\delta)Y(t_0+\delta,t_0-\delta)Y(t_0-\delta,s).
\ee
The first and last  terms of the right hand side can be estimates by
Lemma \ref{easy}, whereas we get for the middle term 
\be
\|Y(t_0+\delta,t_0-\delta)\|\leq e^{\frac{1}{\eps}
\int^{t_0+\delta}_{t_0-\delta}
\|N(u)+\eps A(u)\|du}.
\ee
Altogether this yields
\bea
\|Y(t,s)\|&\leq &
c^2 \|S^{-1}(t)\|\|S(t_0+\delta)\|\|S^{-1}(t_0-\delta)\|\|S(s)\|/{\eps^{2(d-1)/d}}\\ \nonumber
& &\times 
e^{ (\int_{s}^{t_0-\delta}+\int_{t_0+\delta}^t) (1+c\|B(u)\|) 
du/\eps^{(d-1)/d}}e^{ +\frac{1}{\eps}
\int^{t_0+\delta}_{t_0-\delta}
\|N(u)+\eps A(u)\|du}.
\eea
By (\ref{ss}), there exists a constant $c$ (that may change from line to  
line) which in dependent of $\eps$ such that the pre-exponential factors 
are bounded by $c/\delta^{2p_0}$. Also, since $N(t)$ is $C^1$ and $A(t)$ is
$C^0$ on $[0,1]$, 
\be
\int^{t_0+\delta}_{t_0-\delta}{\|N(u)+\eps A(u)\|}\,du\leq c\delta \ \ \mbox{and } \ \
\int_{t_0+\delta}^t \|B(u)\|\,du\leq c/{\delta^{p_0}} 
\ee
and similarly for $\int_s^{t_0-\delta} \|B(u)\|$. Hence, 
$Y(t,s)$ satisfies the bound
\be
\|Y(t,s)\| \leq c e^{c(\frac{1}
{\delta^{p_0}\eps^{(d-1)/d}}+\frac{\delta}{\eps})}/({\delta^{2p_0}}{\eps^{2(d-1)/d}}).
\ee
Choosing $\delta=\delta(\eps)=\eps^{\frac{1}{d(p_0+1)}}$ in order to balance 
the contributions in the 
exponent, we get with a suitable constant $c$
\be
\|Y(t,s)\| \leq c e^{c/\eps^{\frac{(p_0+1)d-1}{(p_0+1)d}}}/
\eps^{\frac{2(d(p_0+1)-1)}{(p_0+1)d}}.
\ee
Picking $\frac{(p_0+1)d-1}{(p_0+1)d}<\beta_0 <1$, we get for 
yet another constant $c$
\be
\|Y(t,s)\| \leq c e^{c/\eps^{\beta_0}}.
\ee
A similar analysis yields the same result in case $t_0=u$ or $t_0=w$.
As there are only finitely many weakly splitting points to take care of,
taking for $\beta<1$ the largest of the $\beta_j$, for $j=1, \cdots,m$, 
we get the result.
\ep

\vspace{.4cm}

\noindent
{\bf Remarks:} \\
i) The proof is valid in arbitrary dimension, 
assuming only (\ref{ss}) at a finite number of points.\\
ii) The exponents $p_i>0$ in (\ref{ss}) need not be integers. 
 
\vspace{.4cm}

\noindent
{\bf Proof of Lemma \ref{compic}:} 
Let $Y(t,s)$ be a solution to
(\ref{evonilp}) and assume $N(u)\equiv 0$ for all $s\leq u \leq t$. Then
$\|Y(t,s)\|\leq e^{\int_s^t\|\ A(u) |\, du}$, which shows one implication. 
We prove the reverse implication by contradiction. Assume there exists $u_0\in [s,t]$ such
that the nilpotent $N(u_0)\neq 0$ and $\|Y(t,s)\|\leq c$, uniformly as $\eps\ra 0$,
for all $0\leq s\leq t\leq 1$. We compare $Y(t,s)$ with
\be
Z_0(t,s)=e^{N(u_0)(t-s)/\eps}
\ee
and get the following estimate from (\ref{difst}) and (\ref{dyson})
\be
\|Z_0(t,s)\|\leq ce^{c\int_s^t \|N(u)-N(u_0)+\eps A(u) \|\, du/\eps}.
\ee
By Taylor's formula, there exists a $\delta>0$ such that $t-s\leq \delta$ implies
\be
\int_s^t \|N(u)-N(u_0)+\eps A(u) \|\, du \leq c\delta(\delta +\eps),
\ee
for another constant $c$. Hence, if $t-s\leq \delta$, with $\delta$ small enough, 
\be\label{bz0}
\|Z_0(t,s)\|\leq ce^{c\delta^2/\eps},
\ee
for some $c$. On the other hand, if $t-s=\delta$ and $\eps <\hspace{-.15cm}< \delta$, we have for some $c$,
\be\label{nz0}
\|Z_0(t,s)\|= c (\delta/\eps)^{d-1}.
\ee
Thus, by letting $\delta$ and $\eps$ tend to zero in such a way that 
$\delta^2 <\hspace{-.15cm}< \eps <\hspace{-.15cm}< \delta$, we get a 
contradiction between (\ref{bz0}) and (\ref{nz0}), which finishes the proof of the 
statement.

\section{Iterative Scheme}\label{is}\setcounter{equation}{0}

We present here the iterative construction which leads to the construction
of $V^{q^*(\eps)}(t,s)$ developed in \cite{jp1}, to which we refer the reader for 
proofs and more details. The first general construction of this kind is to be
found in \cite{n1}. \vspace{.3cm}

Assume {\bf H1} and {\bf H2} with $a>0$ small enough so that {\bf H2} holds
in $S_a$. \\

By perturbation theory in $z\in S_a$, if $z_0\in S_a$ and $\Gamma_j\in 
\rho(H(z_0))$, $j=1, \cdots, n$ are simple loops encircling the eigenvalues 
$\lambda_j(z_0)$, there exists $r>0$ such that for any  $z\in B(z_0,r)$, 
 where $B(z_0,r)$ is an open disc of radius $r$ centered at
$z_0$, $\Gamma_j\in \rho(H(z))$,

For $z\in B(z_0,r)$, we set 
\bea
P_j(z)&=&-\frac{1}{2\pi i}\int_{\Gamma_j}(H(z)-\lambda)^{-1}\, d\lambda\equiv P_j^0(z),
\ \ \  P_0(z)=P_0^0(z),\\
K^0(z)&=&i\sum_{k=0}^n{P_k^{0}}'(z)P_k^{0}(z).
\eea
The operator
$K^0$ is bounded, analytic and we define the closed operator
\be
H^1(z)=H(z)-\eps K^0(z) \ \ \mbox{on} \ \ \mathcal D.
\ee
For $\eps $ small enough, the gap hypothesis {\bf H2} holds for all $z\in B(z_0,r)$,
and we set for  $\eps $ small enough
\bea
P_j^1(z)&=&-\frac{1}{2\pi i}\int_{\Gamma_j}(H^1(z)-\lambda)^{-1}\, d\lambda, 
\ \ P_0^1=\mathbb I -\sum_{j=1}^nP_j^1(z)\\
K^1(z)&=&i\sum_{k=0}^n{P_k^{1}}'(z)P_k^{1}(z).
\eea
Note that $H^1$, $P_k^1$, $k=0,\cdots , n$,  and $K^1$ are $\eps$-dependent and
strongly analytic in $B(z_0,r)$.

We define inductively, for $\eps$ small enough, the following hierachy of operators
for $q\geq 1$
\bea
H^q(z)&=&H(z)-\eps K^{q-1}(z)\\ \label{pq}
P_j^q(z)&=&-\frac{1}{2\pi i}\int_{\Gamma_j}(H^q(z)-\lambda)^{-1}\, d\lambda, 
\ \ P_0^q=\mathbb I -\sum_{j=1}^nP_j^q(z)\\
K^q(z)&=&i\sum_{k=0}^n{P_k^{q}}'(z)P_k^{q}(z).
\eea

It is proven among other things in \cite{jp1}, see also \cite{jp2}, that the 
following holds:
\begin{prop}\label{jpjmp}
There exists $\eps_0>0$, $b>0$ and $g>0$ such that for all 
$q\leq q^*(\eps)\equiv [g/\eps]$ and all $z\in B(z_0,r)$,
$K^q(z)$ is analytic in $S_a$, and 
\bea
&&\|K^{q}(z) -K^{q-1}(z)\|\leq b q!\left(\frac{\eps}{eg}\right)^q\\
&&\|K^q(z)\|\leq b.
\eea
\end{prop}
{\bf Remarks:}\\
i) As a corollary, for 
\be\label{qstar}
q=q^*(\eps)=[g/\eps],
\ee
we get the exponential estimate 
\be\label{key}
\|K^{q^*(\eps)}(z) -K^{q^*(\eps)-1}(z)\|\leq  eb \, e^{-g/\eps}.
\ee
\noindent
ii)
The values of $\eps_0$ and $g$  which determines the exponential decay above only  
depend
on 
$$\sup_{z\in B(z_0,r)\atop \lambda \in \cup_{j=1}^n\Gamma_j }\|(H(z)-\lambda)^{-1}\|
<\infty,$$
see \cite{jp1} for explicit constants. \\
iii) Since $S_a$ is compact, at the expense of decreasing the value of $a$, we can 
assume that proposition \ref{jpjmp} holds for any $z\in S_a$, with uniform constants
$g, \eps_0$ and $b$.
\vspace{.3cm}

Before we go on, let us recall a few facts from perturbation theory 
applied to our setting, that will be needed in the sequel.\\

Assume $q\leq q^*(\eps)$ and let 
$\lambda\in \cup_{j=1}^n \Gamma_j \subset \rho(H(z_0))$ and $z\in B(z_0, r)$. 
We can write for
$\eps<\eps_0$
\bea
(H^q(z)-\lambda)^{-1}&=&(H(z)-\eps K^{q-1}(z)-\lambda)^{-1}
\\ \nonumber&=& (H(z)-\lambda)^{-1} +\eps(H(z)-\lambda)^{-1}K^{q-1}(z) 
(H^q(z)-\lambda)^{-1}
\\ \nonumber
&=&  (\mathbb I -\eps(H(z)-\lambda)^{-1}K^{q-1}(z) )^{-1}(H(z)-\lambda)^{-1} .
\eea
Hence, for any $j=1,\cdots, n$, 
\bea\label{perpj}
P_j^q(z)&=&P_j(z)-\frac{\eps}{2\pi i}\int_{\Gamma_j}(H(z)-\lambda)^{-1}K^{q-1}(z) 
(H^q(z)-\lambda)^{-1} \, d\lambda \nonumber\\
&=&P_j(z)-\eps R_j^q(z)
\eea
is analytic in $z$ and the remainder is of order $\eps$, together with all its 
derivatives. Moreover, making use of 
\be
(H(z)-\lambda)^{-1}=(H(z)-\lambda_0)^{-1}(\mathbb I-
(\lambda-\lambda_0)(H(z)-\lambda)^{-1})
\ee
for $\lambda_0$ in $\rho(H(z))$, we can write
\be
H(z)P_j^q(z)=H(z)P_j(z) + \eps F_j^q(z)
\ee
where $F_j^q(z)$ given by
\be\label{fq}
H(z)(H(z)-\lambda_0)^{-1}
\int_{\Gamma_j}(\mathbb I-
(\lambda-\lambda_0)(H(z)-\lambda)^{-1})K^{q-1}(z) 
(H^q(z)-\lambda)^{-1} \frac{d\lambda}{2\pi i}.
\ee
The identity
\be
H(z)(H(z)-\lambda_0)^{-1}=\mathbb I+\lambda_0(H(z)-\lambda_0)^{-1},
\ee
shows that  $F^q_j(z)$
is uniformly bounded as $\eps\ra 0$ and analytic. 

As a consequence, we have 
\begin{lem}\label{hqpq} Let $F_j^q$ be defined by (\ref{fq}). Then
\bea
H^q(z)P_j^q(z)&=&H(z)P_j(z)+\eps(F_j^q(z)-K^{q-1}(z)P_j^q(z))\\
H^q(z)P_0^q(z)&=&H_0(z)+\eps(F_0^q(z)-K^{q-1}(z)P_0^q(z)),
\eea
where $F_0^q(z)=-\sum_{j=1}^nF_j^q(z)$.
\end{lem}

\section{The Approximation}\label{a}\setcounter{equation}{0}

Let $q\leq q^*(\eps)$ and consider $V^q$, defined as the solution to
\bea\label{vq}
&&i\eps \partial_t V^q(t,s)\ffi=(H^q(t)+\eps K^q(t))V^q(t,s)\ffi, \\ \nonumber 
&& \ffi\in D, \ \ V^q(s,s)=\mathbb I, \ \ 0\leq s \leq t \leq 1.
\eea
As $H^q=H-\eps K^{q-1}$ we get that 
\be
H^q(t)+\eps K^q(t)=H_0(t)+\sum_{j=1}^n P_j(t)H(t)P_j(t)+\eps (K^q(t)-K^{q-1}(t))
\ee
is a bounded, smooth perturbation of $H_0(t)$. The results of \cite{kr} guarantee
the existence and uniqueness of the solution to (\ref{vq}).
Moreover, as is well known \cite{k}, \cite{kr}, $V^q$ further satisfies
\be\label{interq}
V^q(t,s)P^q_k(s)=P^q_k(t)V^q(t,s), \ \ \forall k=0,\cdots, n, \ \ 0\leq s\leq t\leq 1.
\ee

\vspace{.3cm}

In order to show by means of (\ref{pert}) that $V^q$, with $q=q^*(\eps)$, 
is a good approximation of $U$,  we need to control the behaviour of the 
norm of $V^q$ as $\eps \ra 0$. 
We split $V^q$ into components within the spectral subspaces of $P_k^q$.  
Set
\be\label{defcompv}
V_k^q(t,s)=V^q(t,s)P_k^q(s) \ \ \ \mbox{s.t.} \ \ \ V^q(t,s)=\sum_{k=1}^nV_k^q(t,s).
\ee
Since the projectors $\{P_k^q(s)\}_{k=0,\cdots, n}$ have norms uniformy bounded
from above and below in $s\in[0,1]$ and $\eps >0$, there exists a positive constant
$\gamma$ such that
\be\label{eqnorm}
\gamma^{-1} \max_{k=0,\cdots, n}\|V_k^q(t,s) \|\leq \|V^q(t,s) \|\leq 
\gamma \max_{k=0,\cdots, n}\|V_k^q(t,s) \|.
\ee
We have,
\begin{prop}\label{nv}
There exist constants $C_k>0$, $k=0,1, \cdots, n$, $d_j\geq 0$ and $0<\beta_j<1$, 
$j=1,\cdots, n$
such that for all $\eps < \eps_0$, and all $q\leq q^*(\eps)$,
\bea
\|V_0^q(t,s) \|=\|V^q(t,s)P^q_0(s)\| &\leq& C_0 e^{\Im \int_s^t\lambda_0(u)du/\eps}\\
\|V_j^q(t,s) \|=\|V^q(t,s)P^q_j(s)\| &\leq& C_j e^{d_j/\eps^{\beta_j}}
e^{\Im \int_s^t\lambda_j(u)du/\eps}.\label{esdj}
\eea
Moreover, (\ref{esdj}) holds with  $d_j=0$ if and only if $D_j(t)\equiv 0$ in (\ref{sd}).
\end{prop}
\noindent
{\bf Proof of Proposition \ref{nv}:}\\
We first consider $V_0^q(t,s)$, the part of $V^q$ corresponding  to the infinite 
dimensional subspace $P_0^q$. Because of (\ref{interq}), it satisfies 
for  $0\leq s \leq t \leq 1$ and any $\ffi\in D$
\be
i\eps\partial_tV_0^q(t,s)\ffi=((H^q(t)+\eps K^q(t))P_0^q(t))V_0^q(t,s)\ffi, \ \
 V_0^q(s,s)=P_0^q(s).
\ee
Lemma \ref{hqpq} shows that the generator of $V_0^q(t,s)$ is equal to 
$H_0(t)$ plus a smooth bounded perturbation of order $\eps$. We can thus 
compare $V_0^q(t,s)$ and $U(t,s)P_0^q(s)$ by means of (\ref{pert}). The fact
that the initial condition is $P_0^q(s)$ instead of the identity simply
multiplies  the estimate by  $\|P^q_0(s)\|$, so that we get
\be\label{esvq}
\|V^q(t,s)P^q_0(s)\| \leq  \|P^q_0(s)\| e^{\Im \int_s^t\lambda_0(u)du/\eps} C'_0
\leq e^{\Im \int_s^t\lambda_0(u)du/\eps} C_0,
\ee
where $C_0'$ and $C_0=C_0' \sup_{s\in [0,1]\atop \eps>0}\|P^q_0(s)\|$ are 
uniform in $\eps$.

The control of the remaining components is conveniently done by taking advantage of 
the intertwining  relation (\ref{interq}) as follows. 

Let $W^q$ be the bounded operator satisfying the equation
\be\label{kato}
i{W^q}'(t)=K^q(t)W^q(t), \ \ \ W^q(0)=\mathbb I.
\ee
This operator enjoys a certain number of properties.
As $K^q$ is smooth and bounded, the solution is given by a convergent
Dyson series, and $W^q(t)$  interwines between $P^q_k(0)$ and $P^q_k(t)$.
Morerover,  ${W^q}$ and its inverse map $D$ into $D$, see \cite{jp0}. Finally, 
by regular perturbation theory and Proposition \ref{jpjmp}, $K^q=K^0+O(\eps)$ 
so that
\be\label{nwq}
\sup_{t\in [0,1]\atop 0< \eps <1}\|{W^q(t)}^{\pm 1}\|<\infty .
\ee
Therefore, the bounded operator defined by
\be\label{defphi}
\Phi^q(t,s)={W^q(t)}^{-1}V^q(t,s)W^q(s), \ \ \ 0\leq s\leq t\leq 1
\ee
satisfies by construction 
\be
[\Phi^q(t,s), P_k^q(0)] \equiv 0, \ \ \ \forall \ k=0,\cdots, n\ \ \  
\forall \ 0\leq s\leq t\leq 1.
\ee
We can thus view 
\bea
\Phi_j^q(t,s)&=&\Phi^q(t,s)P_j^q(0), \ \ \ j=1,\cdots , n,\\
\Phi_0^q(t,s)&=&\Phi^q(t,s)P_0^q(0)
\eea
as operators in the finite dimensional Banach spaces $P_j(0)\mathcal B$, for $j\geq 1$
and in the infinite dimensional  Banach space $P_0(0)\mathcal B$. Moreover, thanks to
(\ref{nwq}), there exists a constant $C$ such that, uniformly in $0\leq s\leq t\leq 1$ and 
$\eps >0$,
\be\label{vphi}
C^{-1}\|V^q_k(t,s)\|\leq \|\Phi^q_k(t,s)\|\leq C\|V^q_k(t,s)\|, \ \ \ k=0,\cdots, n.
\ee
The operator $\Phi_j^q(t,s)$ satisfies for any $\ffi\in D$
\bea\label{eqphij}
i\eps \partial_t \Phi_j^q(t,s)\ffi&=&
{W^q(t)}^{-1}H^q(t)V^q(t,s)W^q(s)P_j^q(0)\Phi_j^q(t,s)\nonumber\\
&=&P_j^q(0){W^q(t)}^{-1}H^q(t)P_j^q(t)W^q(t))P_j^q(0)\Phi_j^q(t,s)\ffi\nonumber\\
&\equiv& \widetilde H_j^q(t)\Phi_j^q(t,s)\ffi,
\eea
where the generator $ \widetilde H_j^q(t)$ is bounded, see Lemma \ref{hqpq}.
In a sense, $\Phi_j^q(t,s)$ describes the evolution within the spectral
subspaces.  Let us further compute with $P_j^q=(P_j^q)^2$ 
and  (\ref{perpj})
\bea\label{internalh}
 H^q(t)P_j^q(t)&=&P_j^q(t) (H(t)P_j(t)+\eps(F_j^q(t) - K^{q-1}(t))P_j^q(t)\\
&=&P_j^q(t)(\lambda_j(t)P_j(t)+D_j(t)+\eps(F_j^q(t) - K^{q-1}(t) )P_j^q(t)\nonumber\\
&=&\lambda_j(t) P_j^q(t)+P_j^q(t)D_j(t)P_j^q(t)  \nonumber\\
& &+\eps  P_j^q(t)(\lambda_j(t) R_j^q(t) + F_j^q(t) - K^{q-1}(t) ) P_j^q(t)\nonumber\\
\nonumber
&\equiv & P_j^q(t)(\lambda_j(t)+D_j(t))P_j^q(t)+\eps J_j^q(t).
\eea
The last term is bounded, analytic in $t$ and of order $\eps$. We will deal with
it perturbatively.

Equations (\ref{internalh})  suggests to decompose $\Phi_j^q(t,s)$, 
$j=1,\cdots, n$, as
\be\label{defpsi}
\Phi_j^q(t,s)=e^{-i\int_s^t\lambda_j(u)\, du/\eps}\Psi_j^q(t,s)
\ee
where $\Psi_j^q(t,s):P_j^q(0)\mathcal B \ra P_j^q(0)\mathcal B $ satisfies
\bea\label{edpsi}
i\eps \partial_t \Psi_j^q(t,s)&=&P_j^q(0)({W^q(t)}^{-1} (D_j(t) +
\eps J_j^q(t)) W^q(t) P_j^q(0)
\, \Psi_j^q(t,s), \\ \nonumber
 \Psi_j^q(s,s)&=&P_j^q(0) ,
\eea
where, in the leading part of the generator,  
\be
\widetilde D_j(t)= {W^q(t)}^{-1}  D_j(t)W^q(t)
\ee
is analytic and nilpotent with $\widetilde D_j(t)^{m_j}=0$, with $m_j=\dim P_j(t)$.  
However, the restriction of $\widetilde D_j(t)$ to $P_j^q(0)\mathcal B$,
$P_j^q(0)\widetilde D_j(t) P_j^q(0)$, is not nilpotent. Nevertheless, $\Psi_j^q(t,s)$
satisfies the same type of estimates an evolution generated by a perturbed  analytic 
nilponent does:
\begin{lem}\label{nphi}
Let $\Psi_j^q(t,s)$ be defined by (\ref{defpsi}), for $j=1,\cdots, n$. Then, there exist 
$0<\beta_j<1$ and $d_j\geq 0$, $c_j>0$ such that
\be
\|\Psi_j^q(t,s)\|\leq c_j e^{d_j/\eps^{\beta_j}}.
\ee
Moreover, the estimate holds with $d_j=0$ if and only if $D_j(t)\equiv 0$ in (\ref{sd}).
\end{lem}
{\bf Proof of  Lemma \ref{nphi}:} 
Equations  (\ref{perpj}) and (\ref{propeig}) allow to get rid of the projectors 
$ P_j^q(0)$ in (\ref{edpsi}) up to an error of order $\eps$,
\be\label{gotonil}
P_j^q(0)\widetilde D_j(t) P_j^q(0) = 
{W^q(t)}^{-1}P_j^q(t) D_j(t) P_j^q(t)W^q(t)=\widetilde D_j(t)+\eps 
L_j^q(t),
\ee
where 
\bea\label{rl}
L_j^q(t)&=& -{W^q(t)}^{-1} \left(R_j^q(t)D_j(t) P_j(t)+  P_j(t)D_j(t)R_j^q(t)\right. 
\nonumber\\
& &-
\left. \eps R_j^q(t)D_j(t)R_j^q(t) \right) W^q(t)
\eea
is analytic and of order $\eps^0$. Since  $W^q(t)^{\pm 1}$ is analytic and uniformly
bounded, the nilpotent $\widetilde D_j(t)$ satisfies  {\bf N1} uniformly in $\eps>0$,
and  (\ref{edpsi}) and (\ref{rl}) show that the generator of $\Psi_j^q(t,s)$  satisfies 
the hypotheses of Proposition \ref{nilpot}, which yields the estimate. The last statement 
stems from Lemma \ref{compic} . \ep\\

It remains to gather (\ref{vphi}), (\ref{defpsi}) and Lemma \ref{nphi} to end the
proof of Proposition \ref{nv}. \ep

\subsection{End of the Proof}\label{eop}

Given Proposition \ref{nv}, we can finish the proof of our main statement as 
follows. 

Applying (\ref{difst}) to $U$ and $V^q$, we get
\be\label{difuv}
U(t,r)=V^q(t,r)+i\int_r^t V^q(t,s)(K^q(s)-K^{q-1}(s))U(s,r)  \,  ds.
\ee
Let $t\mapsto \omega(t)$ be the continuous function defined by
\be\label{defom}
\omega(t)=\max_{k=0,\cdots, n}\Im \lambda_{k}(t).
\ee
Applying (\ref{difuv}) to $P_k^q(r)$ and multiplication  
by $e^{-\int_r^t\, \omega(s)\, ds/\eps}$ gives with (\ref{defcompv})
\bea
&&\|e^{-\int_r^t \omega(u)\, du/\eps}(U(t,r)-V^q(t,r))P_k^q(r)\|\leq 
\int_r^t \|e^{-\int_s^t \omega(u)\, du/\eps}V^q(t,s)(K^q(s)-K^{q-1}(s))\| \nonumber\\
&&\qquad\times \left(\|e^{-\int_r^s \omega(u)\, du/\eps} 
(U(s,r)-V^q(s,r))P_k^q(r)\| + \|e^{-\int_r^s \omega(u)\, du/\eps} V^q_k(s,r)\|
 \right)\,  ds.
\eea
Proposition \ref{nv} and the definition of $\omega(t)$ yield
for any $0\leq r\leq s\leq 1$
\be\label{fures}
\|e^{-\int_r^s \omega(u)\, du/\eps} V^q_k(s,r)\|\leq C_k e^{d_k/\eps^{\beta_k}}, 
\ \ \ \ \mbox{(with $d_0=0$).}
\ee
Further taking $q=q^*(\eps)$,
(\ref{key}), (\ref{eqnorm}) show the existence of constants $B>0$ and $0<\kappa<g$ 
such that
\bea
&&\|e^{-\int_s^t \omega(u)\, du/\eps}V^{q^*(\eps)}(t,s)(K^{q^*(\eps)}(s)-
K^{q^*(\eps)-1}(s))\|\leq
ebC e^{D/\eps^{\beta}}e^{-g/\eps}\leq B e^{-\kappa /\eps}.
\eea
Hence, we get using $0\leq t-s\leq 1$,
\bea
&&\|e^{-\int_r^t \omega(u)\, du/\eps}(U(t,r)-V^{q^*(\eps)}(t,r))P_k^{q^*(\eps)}(r)\|\\
&&\qquad \qquad\qquad \qquad \leq 
B e^{-\kappa /\eps}\int_r^t
\|e^{-\int_r^s \omega(u)\, du/\eps} V^{q^*(\eps)}_k(s,r)\|\, ds\nonumber\\ \nonumber
&&\qquad \qquad\qquad \qquad  +B e^{-\kappa /\eps}\sup_{ r \leq s \leq t }\|
e^{-\int_r^s \omega(u)\, du/\eps}(U(s,r)-V^{q^*(\eps)}(s,r))P_k^{q^*(\eps)}(r)\|, 
\eea
from which we deduce that if $\eps$ is so small that $B e^{-\kappa /\eps}<1/2$,
\bea
&&\sup_{r  \leq s \leq t}\|e^{-\int_r^s \omega(u)\, du/\eps}(U(s,r)-V^{q^*(\eps)}(s,r))
P_k^{q^*(\eps)}(r)\|
\nonumber\\ && \qquad \qquad \qquad \qquad \leq  2B e^{-\kappa /\eps}(t-r)
\sup_{ r  \leq s \leq t}\|e^{-\int_r^s \omega(u)\, du/\eps} V^{q^*(\eps)}_k(s,r)\|.
\eea
In particular, our main result  follows. For $\eps$ small enough, for
any $0\leq r\leq t\leq 1$, and for all $k=0,\cdots, n$, 
\bea\label{maines}
 e^{-\int_r^t \omega(u)\, du/\eps}U(t,s)P_k^{q^*(\eps)}(r)&=&
e^{-\int_r^t \omega(u)\, du/\eps} 
V^{q^*(\eps)}_k(t,r)\\
\nonumber
& &+O((t-r) e^{-\kappa /\eps}\sup_{ r  \leq s \leq t}\|e^{-\int_r^s \omega(u)\, du/\eps} 
V^{q^*(\eps)}_k(s,r)\|).
\eea
We chose to estimate the difference $U-V^{q^*(\eps)}$ applied on the 
projectors, because the norms of the different components $V^{q^*(\eps)}_k$
vary with $k$. Of course, (\ref{maines}) also holds with $P_k^{q^*(\eps)}$ removed and
$V^{q^*(\eps)}$ in place of $V^{q^*(\eps)}_k$.

Making further use of (\ref{fures}) in the error term of (\ref{maines}), we get 
(lowering the value of $0<\kappa<g$)
\be
U(t,r)=V^{q^*(\eps)}(t,r)+
O((t-r)e^{-\kappa /\eps}e^{\int_r^t \omega(u)\, du/\eps}),
\ee
where $V^{q^*(\eps)}(t,r)=O(e^{\int_r^t \omega(u)\, du/\eps}e^{D/\eps^{\beta}})$,
for some $0<\beta<1$, and $D\geq 0$.
\ep

\end{document}